\documentstyle[preprint,aps,epsf]{revtex}
\begin{document}
\draft
\title{Large positive magnetic susceptibility of nanotube torus}
\author{Ryo Tamura}
\address{Faculty of engineering, Shizuoka University, Johoku 3-5-1,
  Hamamatsu 432-8561, Japan}
\author{Mitsuhiro Ikuta, Toru Hirahara and Masaru Tsukada }
\address{Department of Physics, Graduate School of Science, University
  of Tokyo, Hongo 7-3-1, Bunkyo-ku, Tokyo 113, Japan}
\maketitle

\begin{abstract}
 We calculate the magnetic moment
 caused by the persistent currents 
 in polygonal 
 carbon nanotube tori  by the tight binding model.
 The polygonal CNT tori are formed
 by introducing heptagonal
 and pentagonal defects along the inner hole and outer fringe,
 respectively.
 We found a new
 type of large paramagnetic
 persistent current  caused by the semi-metallic band
 structures of the periodic CNT junction.
 It is  contrasted with the persistent current
 originated from the metallic band structure.

 In spite of diamagnetism of the graphite,
 this paramagnetism is caused by
 not only the magnetic flux in the inner
 hole (AB flux)
 but also
 the magnetic flux directly penetrating the graphite plane
 (direct flux).
 This magnetic moment
 is close to that calculated with only Aharonov Bohm
 effect where  the direct flux  is included
 in the AB flux.

\end{abstract}
 
\section{ Introduction }
In spite of semi-metallic nature of the graphite,
 the carbon nanotube (CNT) 
 becomes metallic and semi-conducting depending
 on the radius and the helicity of  the honeycomb lattice.
 \cite{tubeHamada}
 This promising feature in  nanotechnology
 is caused by
  the quantum size effect;
 the system  smaller than
 the coherent length 
 shows  electronic states  different from
  those of the  macroscopic system.
Another example 
 of the quantum size effect can be seen
 in the persistent current of the mesoscopic
 ring.
An ideal clean mesoscopic one-dimensional 
 metal ring is
 diamagnetic when the number of the electron 
is $4n+2$ with an integer $n$, while
 it is paramagnetic otherwise.\cite{theory-Loss} 
It means that the ring can have opposite sign of the magnetic
 susceptibility compared to the bulk system.
We should notice that this positive magnetic susceptibility
 is inversely proportional to
 the temperature, i.e., it becomes divergent at the absolute zero
 temperature.\cite{theory-Loss,theory-Cheung,theory-Lin} 
 Though it seems similar to the Curie Wise rule,
 the origin is not the spin but the persistent current.
 As there is no general terminology for this paramagnetism,
 we call it giant orbital paramagnetism (GOP) in the present 
 paper.

We are interested in the CNT ring discovered recently
 for the following reasons. \cite{NTring-exp} 
 As for  the magnetic susceptibility 
 with respect to the magnetic field perpendicular to
 the honeycomb plane,
  the CNTs  and the graphite show similar negative values,
 \cite{Ajiki-mag}
 whereas
  the persistent current of the CNT ring can show the GOP.
 \cite{theory-Lin}
Furthermore 
in order to make the radius of the ring $R$  
 nanometer-sized,
 the CNT ring will be more suitable 
 than the conventional mesoscopic ring of metals or GaAs/AlGaAS.
 \cite{persistent-exp} 
 Motivations to downsize the persistent current ring are
 as follows;
(1)The circulating persistent current $I$
 equals $ev_F/(2\pi R)$ with Fermi velocity $v_F$,
 and is related to the magnetic moment per ring $M$  
 as $M=I\pi R^2$. 
 In the rings closed-packed, therefore,
 the magnetic moment per unit area 
 is proportional to $I$ and 
 can be increased by shrinking $R$.
 (2) The persistent current occurs on condition that
 the coherent length $l_{\phi}$ is larger than $ 2\pi R$.
 Since $l_{\phi}$ decreases as the temperature rises,
 the persistent current can be observed at higher
 temperature as $R$ decreases.

 In other theoretical works
 on the persistent current of the CNT ring,
 both  the 'disclination' and  the 'direct flux'
 are not considered,\cite{theory-Lin,Latil,Latge,Meunier}
 whereas we consider both of them.
 Shrinking the radius of the CNT ring $R$ 
 increases  importance of these two factors.
 Firstly it is explained for the disclination as follows.
 Figure 1 illustrates a 'polygonal CNT torus'
 where pentagonal and heptagonal defects  
  form the corner of the polygonal shape
 along the outer fringe and the inner fringe, respectively.
 These defects are generally called disclinations
 and have  significant effect on 
 the CNT junction,\cite{junction} the CNT cap \cite{cap} and the helical CNT.\cite{helixAkagiprb,kphelixTamura}
 Here we define a 'polygonal' CNT torus
 as the CNT torus with the disclinations,
 though it can also have a rounded
 shape when  the disclinations  are close to each other.
\cite{torusIhara}
 Only few attempts have so far been made at
 the persistent current of the polygonal CNT
 torus \cite{Haddon}  in contrast to  that of 
 the 'circular CNT tori'
 discussed by many authors. \cite{theory-Lin,Latil,Latge,Meunier}
 Here we define a 'circular' CNT torus as 
 the CNT torus  formed 
 by elastic deformation of the straight CNT 
 with no disclination.
 In the circular CNT tori, the honeycomb lattice
 shrinks  along the tube axis
 with a factor $R_1/(R_1+D)$ 
 at the inner fringe
 compared to that at the outer fringe.
 Here $R_1$ and $D$ denote the radius
 of the inner hole and the diameter of the CNT, respectively.
 Though this axial strain is relaxed 
 to a certain  degree
 by   the elastic deformation called
 'buckling',\cite{buckling}
 it  will not be enough
 to keep the stability if $R_1$ is close to  $D$.
 In Fig.1,
 on the other hand, 
 this strain is relaxed
 by decreasing the number of hexagons 
 along the inner hole side.
 Though it is widely known that 
 the impurity reduces the GOP,\cite{theory-Cheung,Latil}
 the effect of the disclination on the GOP is
 open to question.\cite{theory-Cheung,theory-Lin,Latil,Latge,Meunier}
 Secondly, the importance of the direct flux is explained
 as follows. We consider both
 the magnetic flux in the graphite plane
 and that in the inner hole
 while  other researchers have investigated 
 only  the latter  flux.
 We call the former flux 'direct flux'
 and  the latter flux 'AB flux' 
 because the former
 touches the electron directly
 and the latter induces the Aharonov-Bohm effect.
 Ratio of the AB flux to the direct
 flux is $R_1^2$ to $(R_1+D)^2-R_1^2$.
 This ratio indicates that the direct flux cannot be neglected 
 when $R_1$ is comparable to $D$.
 Because the graphite is diamagnetic under the direct flux,
 we have to find out whether the direct flux reduces
 the GOP.
 In order to clarify these open questions,
 we discuss the influence of
 the disclination and the direct flux
 on  the magnetic moment of the polygonal CNT torus.

\section{Tight binding model}
 When the polygonal CNT torus is composed
 of only semi-conducting CNTs,
 it has finite HOMO-LUMO gap.
 As can be known later, it means that
 the GOP does not appear.
 Thus our discussion
 can be limited to the metallic CNTs.
 As first examples,
 we discuss the polygonal CNT torus  with symmetry D$_{6h}$
 formed out of the armchair CNTs which are metallic
 irrespective of their radii.
 We use the tight binding (TB) model 
with only $\pi$ orbital.
The position of the atom is given by 
 the assumed polygonal shape of the torus.

Figure 2  illustrates a part of the projection map
 of a polygonal CNT torus.
 The torus is composed of six unit cells, one of which
 is represented by the  rectangle ABB$'$A$'$.
The prime $'$ means that points X$'$ and X become identical
  when actual
 three dimensional shape is formed.
 Vector  $ \vec {L} $ denotes
 the circumference of the original armchair CNT.
The length and direction
 of the CNT axis per unit cell along
 the outer fringe (inner fringe) 
 is represented by  $\vec{ S}$ ($\vec S_2$).
The $i$ membered ring disclination 
 is formed at P$_i$ and Q$_i$ $(i=5,7)$
 by removing the shaded area 
 and sticking  the lines
 P$_5$P$_7$, P$_7$D$'$, DQ$_7$
and Q$_7$Q$_5$ on the lines P$_5$P$'$$_7$,
P$'$$_7$C$'$, CQ$'$$_7$ and Q$'$$_7$Q$_5$, respectively.
 The disclinations
  approximately keep the sp$^2$ bonds of carbon atoms,
 i.e., the number of nearest neighbors is three for each atom.
 In spite of  the unchanged local structure,
 the phase of the wave function is shifted
 when the electron circulate around the disclination
 owing to excess or  deficiency of the $\pi/3$
 angle on the projection map.
 This topological 
 effect cannot be represented by the effective potential
 energy or modification of the bond strength.

Because of the D$_{6h}$ symmetry,
$\vec R_{55}$ and  $\vec R_{77}$ are parallel to $\vec L$, while
   $\vec S_2$ and $\vec S$ are perpendicular to $\vec L$, where 
 the definition of $\vec R_{i,j}$ is clearly illustrated in Fig.2.
 Thus the four integer parameters,
 $n_L=|\vec L|/(\sqrt{3}a)$,  $n_{75}=|\vec R_{75}|/a$, 
 $n_S=|\vec S|/a$
 and $n_{77}=|\vec R_{77}|/(\sqrt{3}a)$ specify the CNT torus
 with $a\simeq$0.25 nm being lattice
 constant of the graphite.
 For example, 
 these parameters are 
 $n_L=6, n_S=5,n_{75}=2$ and $n_{77}=2$ in Fig.2.
 To keep the bond lengths almost constant,
 the cross section cut along
 P$_5$P$_7$Q$_7$Q$_5$P$_5$  has to be rectangular,
 so the four integers have to satisfy the condition 
 $n_{77}=(n_L-n_{75})/2$.
 We choose  the uniform magnetic field
 $\vec{B}$ parallel to the six-fold 
 rotational axis of the torus.
 In Fig.2, $B_{\bot}$  denotes
 the component of $\vec{B}$ perpendicular to the projection map.

The unit cells are numbered
 along the CNT axis with $n_1$ as shown in Fig.1.
 Figure 2 shows  labels of atoms 
in each unit cell, $(n_2,n_3)$,  where
$n_2$ specifies the zigzag rows along the tube axis.
 At the inner hole side, $n_2=1$
 ($n_2=-1$), and $n_2$ increases (decreases) 
one by one as the zigzag row
  climbs up (down) the inner hole wall and approaches 
 outer fringe. 
In  $n_2$ zigzag row,
 atoms are numbered
 along the CNT axis as $n_3=1,2,\cdots,q(n_2)$.
 For example, $q(1)=q(2)=6,q(3)=7$ in Fig.2.
 Between the neighboring unit cells,
 atoms $(n_1,n_2,1)$ and $(n_1,n_2,q(n_2))$ connect 
 with atoms $(n_1-1,n_2,q(n_2))$ and $(n_1+1,n_2,1)$, respectively.
 With this label,
 matrix elements of Hamiltonian between atoms $n$ and $m$
  are represented by
\begin{equation}
H(B)_{n,m}=-t_{n,m}\exp(i\beta(n,m)B)\;\;.
\label{Hamil0}
\end{equation}
Each atom $n$ has a constant hopping integral
$-t_{n,m}= -t (\sim -3$eV) with the three nearest neighbors $m$
 and $t_{n,m}=0$ with all the other atoms $m$.
The effect of the magnetic fields is included by
 Peierls phases
 $\beta(n,m)B$, \cite{Ajiki} where
\begin{eqnarray}
\beta(n,m) &=& \gamma_{D}(n,m)C_1 +\gamma_{AB}(n,m)C_2
\label{alpha} \\
 C_1&=&\sqrt{3}\pi a^2/(2\phi_0)
 \;\;, \nonumber \\
  C_2&=& \sqrt{3}\pi |\vec S_2|^2/(2\phi_0)\;\;. \nonumber \\
\gamma_{AB}(n,m) &=&
\delta_{n_2,m_2}(
\delta_{n_1,m_1+1}\delta_{n_3,1}\delta_{m_3,q(n_2)}
- \delta_{n_1,m_1-1}\delta_{m_3,1}\delta_{n_3,q(n_2)})
\;\;,\nonumber
 \\
\gamma_{D}(n,m) &= & 
\delta_{n_2,m_2}\{
f(n_2)(\delta_{q(n_2)n_1+n_3,q(n_2)m_1+m_3+1}
\nonumber \\
 & & - \delta_{q(n_2)n_1+n_3,q(n_2)m_1+m_3-1})
+g(n_2)\gamma_{AB}(n,m)  \}
\;\;,\nonumber
\end{eqnarray}
 with the magnetic flux quantum $\phi_0=h/e$
 and  Kronecker delta $\delta$.
Here $f$ and $g$ 
are linear  or constant as a function of $n_2$;

\begin{equation}
f(n_2)=
 \left \{ \begin{array}{ll}
n_{75}&\;\;\cdots   |n_2| \geq 1+ n_{75}+n_{77} \\
|n_2|-n_{77} -0.5 &\;\;\cdots n_{77}+1 \leq |n_2| \leq n_{75}+n_{77} \\
 0&\;\;\cdots |n_2| \leq n_{77} 
\end{array}
\right .
\end{equation}

\begin{equation}
g(n_2)=
 \left \{ \begin{array}{ll}
0 &\;\;\cdots   |n_2| \geq 1+ n_{75}+n_{77} \\
f(n_2)+2\sum_{j=|n_2|+1}^{n_{75}+n_{77}} f(j) &\;\;\cdots |n_2| \leq n_{75}+n_{77}. 
\end{array}
\right .
\end{equation}
 In eq.(\ref{alpha}), $\gamma_{D}(i;j)C_1$ and 
$\gamma_{AB}(i;j)C_2$ come from 
the direct flux and the AB flux, respectively.
Equation (\ref{alpha}) satisfies the condition that
 $\phi_0 B\sum_{i=1}^{j-1}\beta(n^{(i+1)};n^{(i)})/(2\pi)$ 
equals the magnetic flux
 surrounded by the closed loop
$n^{(1)} \rightarrow n^{(2)} \rightarrow \cdots n^{(j)}=n^{(1)}$.

\section{Result}
Firstly we explain the origin of the
 GOP of the circular CNT torus
 with  the dispersion relation $E_l(k,B)$ of the straight CNT
 under the uniform magnetic field $B$  perpendicular to
 the tube axis.
 Ajiki and Ando showed that
 the direct flux flattens the dispersion lines
 near the Fermi level.\cite{Ajiki}
 This is schematically shown by the dashed curves $E_l(k,B)$ 
  compared to the linear dispersion $E_l(k,0)$ in Fig.3. 
When the lattice distortion is neglected,
the energy level of the circular CNT torus
 $\epsilon_{l,j}$  can be expressed by
\begin{equation}
\epsilon_{l,j}=E_l(k_j(B),B) 
\end{equation}
\begin{equation}
k_j(B) = \{j+(BS_{\rm AB}/\phi_0)\}/R
\label{kj}
\end{equation}
 with the area of
 the inner hole $S_{\rm AB}$ and  an integer $j$.
 Equation (\ref{kj}) represents the AB effect
 while difference between $E_l(k_j(B),B)$ and $E_l(k_j(B),0)$
 comes from the direct flux.
Then the total energy $U(B)$ is
 obtained as
 \begin{equation}
U(B)=2\sum_{l,j}
f_{\rm{FD}}(\epsilon_{l,j})\epsilon_{l,j}
\label{total-energy}
\end{equation}
where $f_{{\rm FD}}$ is Fermi-Dirac distribution function
 and factor 2 represents spin degeneracy.
 Zeeman effect is neglected here but
  will be discussed latter.
We consider the case of the absolute zero temperature,
 so $\sum_{l,j}f_{{\rm FD}}$ is replaced by
 the summation $\sum_{l,j =occ}$ limited to the occupied states.

The magnetic moment per torus $M$ is calculated
 as $M=-dU/dB$.
 In order to see the effect of the direct flux,
 we  firstly discuss  
 the magnetic moment induced by only AB flux 
\begin{equation}
M_{\rm AB}=-S_{\rm AB}/(R\phi_0)\sum_{l,j=occ}(dE_l(k,0)/dk)|_{k=k_j(0)}
\;\;.
\label{MAB1}
\end{equation}
 Since $E_l(k,0)$ is an even function of $k$,
  the energy level  at
 $k=k_j(0)$ and that at $k=k_{-j}(0)=-k_j(0)$ 
 are  occupied at the same time.
Their contribution to $M_{\rm AB}$ usually cancels each other,
 because $dE_l(k,0)/dk$ is an odd function of $k$.
 When
 the highest occupied level (HOL) comes at K and K' corner points,
 however,  this cancellation does not occur so that the GOP
 is induced.
 This is illustrated in Fig.3 where
 the energy levels and its change caused by the AB effect
 are indicated by circles and arrows, respectively.
Without the direct flux, the levels move along
 the linear dispersion line and
 only the level  with negative $dE/dk$ is
 occupied (closed circle) while that with positive $dE/dk$ 
 becomes vacant (open circle). 
 Only the former contributes to the magnetic moment
  and brings about the giant orbital paramagnetism (GOP).
 The occupation of only the state with a negative 
$dE/dk$ corresponds to the generation of the persistent current.
The essential point is 
 that the corner point is  {\it partially} occupied (gray circle).
 If the corner point were {\it fully} occupied,
  the open circle  level would be also occupied 
 and it would cancel the energy decrease of the closed circle level.
With the direct flux, however, the closed circle level
  is located on the dashed curve and shifted upward
 compared to the case without the direct flux, i.e.,
 the direct flux  reduces the GOP. 

 The magnetic moment for the polygonal torus can be calculated
 almost the same way as the circular torus.
 In this case, we use the phase $\alpha$  instead of the
 crystal wave number $k$.
 Rotation by $\pi/3$ with respect to the six-fold symmetry
 axis is equivalent to multiplying the wave function
 by the phase factor $\exp (i \alpha)$ as seen in Fig.1.
 Though $k\pi R/3$ corresponds to $\alpha$,
 it should be noted that 
 $R$ cannot be defined uniquely
  for the polygonal torus.
   The energy levels $E_l(\alpha)$ are the eigen values
 of the matrix 
 $H_1(B)+\exp(i\alpha)H_2(B)+\exp(-i\alpha)H_2^{\dag}(B)$ where
 $H_1$ and $H_2$ refer to the bonds in the unit cell
 and those connecting the neighboring unit cells, respectively.
 They are obtained from the Hamiltonian matrix $H$ as
 $H_1(n_2,n_3;m_2,m_3) \equiv H(n_1,n_2,n_3;n_1,m_2,m_3)$,
 and  $H_2(n_2,n_3;m_2,m_3) \equiv H(n_1,n_2,n_3;n_1+1,m_2,m_3)$
 with the label of atoms $(n_1,n_2,n_3)$ defined in section II.
 Because the wave function is invariant under 
 $2\pi$ rotation, 
 $\alpha$ takes discrete values
 $\alpha_j=(\pi/3)(j+(BS_{\rm AB}/\phi_0))$ $(j=$ integer)
 including the AB flux, $BS_{\rm AB}=B 3\sqrt{3}|\vec S_2|^2/2$.
 It results in  discrete levels  $E_l(\alpha_j,B)$
 of the CNT torus.

 On the other hand,
 $E_l(\alpha,B)$ with continuous $\alpha$ 
 represents the continuous energy spectra
 of the  periodic CNT junction made by connecting the  unit
 cell of the CNT torus.
 In the present case,
 the band structure  $E_l(k,0)$
 is classified  by three parameters,
 $n_S/3, n_{75}/3$ and $(n_S-n_{75})/3$
 as listed in Table I
 according to Ref.\cite{helixAkagiprb}.
As shown in Fig.2, the CNT torus contains
two kinds of CNTs.
 One is the original armchair CNT with the
 chiral vector $\vec L$,
 and the other is characterized by different chiral vector $\vec L'$.
When $n_{75}/3$  is an integer, 
 latter  becomes also metallic 
 and we proved analytically
 that types 1 and 2 are semi-conducting
 and metallic, respectively.\cite{kphelixTamura}
 Though we obtained only numerical results
 for the other types,
 we can say that they tend to be semi-conducting as the axial length 
of the semi-conducting CNT segment 
$|\vec L' \times \vec R_{77}|/|\vec L'|$ becomes longer.
 \cite{note}
\begin{table}
\caption{}
\begin{tabular}{|c|c|c|c|c|}
\hline
\multicolumn{5}{|c|}
{$n_{75}/3$} \\
\multicolumn{2}{|c|}{integer}&
\multicolumn{3}{c|}{non-integer}\\
\hline
\multicolumn{2}{|c|}{$n_S/3$}&
\multicolumn{3}{c|}{$n_S/3$}\\
\multicolumn{1}{|c|}{integer}&
\multicolumn{1}{c|}{non-integer}&
\multicolumn{1}{c|}{integer}&
\multicolumn{2}{c|}{non-integer}\\
\hline
 & & & \multicolumn{2}{c|}{$(n_S-2n_{75})/3$ }\\
& & & integer & non-integer \\
\hline
type 1 & type 2 & type 3 & type 4-1 & type 4-2 \\
\hline
semi-conductor &metal &semi-metal & semi-metal & semi-conductor \\
 & & semi-conductor & semi-conductor &  \\
 & & metal & metal &  \\
\hline
\end{tabular}
\end{table}

The dispersion curves
 are shown
 for type 2 periodic CNT junction (metallic) in Fig.4
 and for type 3 periodic CNT junction (semi-metallic) in Fig.5.
 The index $l$ of $E_l(\alpha,B)$ 
 is defined as
 $E_1(\alpha,B) \leq E_2(\alpha,B) \leq \cdots, 
\leq E_{2N}(\alpha,B)$,
 where $2N$ is
 the number of atoms in the unit cell,
 i.e., $N \equiv 2(n_S n_L-n_{57}n_{77})-n_{57}^2$.
Thus the HOMO band is denoted by $E_N(\alpha,B)$ while
 $E_{N+1}(\alpha,B)$ corresponds to the LUMO band.
The circles, lines and arrows in Fig.4 and Fig.5
 have same meanings with those in Fig.3.
 Firstly we consider the magnetic moment under only AB flux $M_{AB}$.
 It is analogous to eq.(\ref{MAB1}) as
\begin{equation}
M_{\rm AB}=-C_2\sum_{l,j=occ}(dE_l(\alpha,0)/d\alpha)|_{\alpha=\alpha_j}
\;\;
\end{equation}
 with $C_2$ defined in eq.(\ref{alpha}).
 The HOMO and LUMO bands  of type 2   is similar to
 those of metallic CNTs,
 but the band crossing point
 is slightly shifted from $\alpha=\pm 2\pi/3$
owing to the phase shift at
 the disclinations. \cite{kphelixTamura}
  Therefore the HOL is fully occupied
 and the GOP does not occur.
Under finite $B$, however,
   $\alpha_{2}=2\pi/3+C_2B$ comes 
 at the cross point so that
  $M_{AB}$ shows a discrete change from negative  to
 positive  as shown by  Fig.6.
 Contrary to it, 
 HOL can be partially occupied
 in the CNT torus of type 3 and 4-1,
 on the condition that  the corresponding  
 periodic CNT junction is semi-metallic.
Figure 5  is an example of it
 where $E_{N}$ at $\alpha=0$  is vacant and
 $E_{N+1}$ at $\alpha=\pm \pi/3$ 
 are  partially occupied instead.
The AB flux  lifts the
 degeneracy of Hogs and only the lowered level at
 $\alpha=\pi/3$ is occupied causing the GOP.
Then  does the direct flux reduce the GOP ?
Surprisingly, it {\it enhances} the GOP.
The dashed lines in Fig.5
 represent
  the dispersion relation
 $E_l(\alpha,B)$ under the finite magnetic
 field $B=0.01\phi_0/a^2$.
In contrast to Fig.3,
the dispersion lines is shifted along the
$\alpha$ axis rather than  along the
$E$ axis.
By this horizontal shift,
the effect of the AB shift
 is not canceled but enhanced.

Here we should notice
 the competition between
 the GOP and spin paramagnetism.
When  the Zeeman  split $g \mu_B B$ ($\mu_B=$ 
 Bohr magneton, $g \simeq 2$) is larger than
  the GOP split $|MB|=|BdE/dB|$ ,
   both the HOLs with positive $dE/dk$ and negative $dE/dk$
 become occupied by the same spin
 so that  spin magnetic moment, that is  $2g \mu_B$   per torus,
 appears 
 instead of the GOP.
 Since $\mu_B \simeq ta^2/\phi_0$ ($t \simeq$ 3 eV,
 $a\simeq 0.25$ nm), the GOP is relevant when
 $M$ is larger than $4t/(\phi_0/a^2)$.
To search the polygonal CNT torus showing the relevant GOP,
the magnetic moment $M(1.5\Delta B)=-(U(2\Delta B)-U(\Delta B))/ \Delta B$ with $\Delta B=0.5\times 10^{-5}\phi_0/a^2$, i.e., 
 $B=1.5\Delta B\simeq 0.24$ Tesla,
 is calculated for  two hundred seven kinds of the tori
 in the range of the parameters $3 \leq n_S \leq 12$,
 $1 \leq n_{75} \leq 6$, $3 \leq  n_L \leq 12$
 and $1 \leq n_{77}= (n_L-n_{75})/2 \leq 6$.
Among the two hundred seven tori,
 the number of the tori with $M$ larger than
 $4ta^2/\phi_0$ is twenty one.
 All of them belong to type 3 or type 4-1
 and   show the GOP.
 Especially  three CNT tori show $M$ larger than
 $ 20 ta^2/\phi_0$.
 In contrast to it, $M$ is much
 less than $4ta^2/\phi_0$ for the tori without the GOP.
 Since $MB$ is approximately the split between the degenerate HOL
 induced by $B$,
 temperature has to be lower than $MB$.
 When $M = 20 t/(\phi_0/a^2)$ and $B=1.5\Delta B$, 
 $MB \sim$ 6 K, which is achievable in the experiment.

 To show the distribution of $M$ and 
 the enhancement of $M$  by the direct flux,
  $M_{AB}/M$ is shown in Fig.7 as a function of $M$
 for the tori showing the relevant GOP.
 For all the tori, $M_{AB} < M$, i.e., the direct flux
 enhances the GOP.
 Figure 7 also shows that
$(|\vec S|^2/|\vec S_2|^2)M_{AB}/M$
 becomes closer to unity than $M_{AB}/M$ where 
the factor $|\vec S|^2/|\vec S_2|^2$ represents
 the ratio of the AB flux 
 to the full flux.
This suggests that the effect of the direct flux
 can be almost included by the AB effect for which
 the cross section is not $3\sqrt{3}|\vec S_2|^2/2$ 
 but $3\sqrt{3}|\vec S|^2/2$.
 It is equivalent to
  replacing $C_1$ and $C_2$ in eq.(\ref{alpha})
 with $ C_1=0$ and $C_2=\sqrt{3}\pi |\vec S|^2/(2\phi_0)$.
In other word, the HOMO and LUMO dispersion curve
can be approximated by
 \begin{equation}
E_l(\alpha,B) \simeq E_l(\alpha+(\pi/3)(BS_D/\phi_0),0)
\label{alphashift}
 \end{equation}
 with the cross section
 of the graphite plane $S_D=3\sqrt{3}(|\vec S|^2-|\vec S_2|^2)/2$.
 In fact, the shift of $\alpha$ in eq.(\ref{alphashift})
 can be seen  in Fig.4  and Fig.5 where
 the dashed lines represent
 the dispersion relation $E_l(\alpha,B=0.01\phi_0/a^2)$.
 It is also the effect of eq.(\ref{alphashift})
 that
 the magnetic moment
 $M$ 
 shows discrete change at 
 lower $B$ than  $M_{\rm AB}$
  in Fig.6.

\section{Summary and Discussion}
In some of the polygonal CNT tori,
 the persistent current
  leads to
 large positive magnetic
  susceptibility
 inversely proportional to the temperature,
 which we call the giant orbital paramagnetism (GOP).
 The magnetic moment per torus can be larger
 than Bohr magneton
 in the achievable condition, $B \sim 0.24$ Tesla  and $T<$ 6 K.
 The necessary condition for the GOP
 is that the corresponding periodic 
 CNT junction is semi-metallic, i.e., type 3 or type 4-1.
 The magnetic flux is divided into
 two parts which are the AB flux in the inner hole
 and the direct flux intersecting the graphite plane.
 In spite of the diamagnetism of the graphite,
 the direct flux enhances the GOP and its effect
 can be effectively included in the AB effect,
 where  the cross section of the AB flux is not
 the inner hole area, but the area surrounded
  by the outer fringe,
 as shown in eq. (\ref{alphashift}) and Fig.7.

 The GOP of the circular CNT torus is caused 
 by the metallic band structure of the straight CNT,
 whereas  that of the polygonal CNT torus
 originates from the semi-metallic band structures
 of the periodic CNT junction.
 This difference comes from change of the bond network,
 i.e.,  removing
 the shaded areas and sticking their
 edges in the projection map illustrated by Fig.2.
 Compared to it, 
 the elastic deformation does not affect
 whether $n$ and $m$ are bonded or not,
 so it is  irrelevant to the GOP of the polygonal CNT torus.
 Owing to this new kind of the GOP, 
 the polygonal CNT torus is a promising nano-structure
 to realize the large persistent current.

\section{Acknowledgment}
 This work was supported in part by a Grant-in-Aid for Creative Scientific
Research on "Devices on molecular and DNA levels" (No. 13GS0017) from the Japan
Society for the Promotion of Science.

\begin{figure}
    \epsfxsize=\columnwidth
\centerline{\hbox{
\epsffile{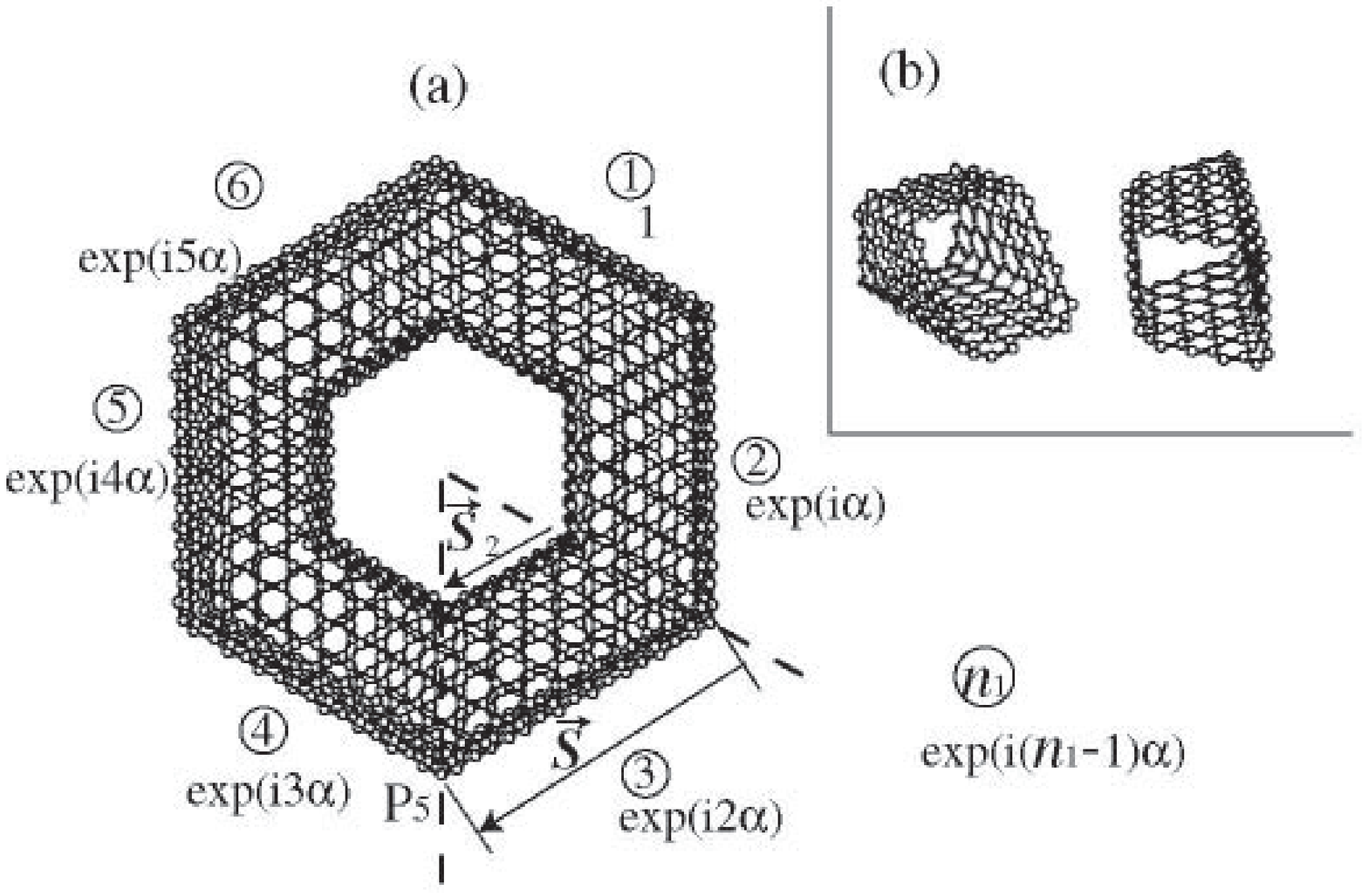}  }}

\caption{(a)Three dimensional shape of the polygonal nanotube torus.
 The applied uniform magnetic field is perpendicular to this page.
 Labels of unit cells $n_1$ are shown.
(b) The unit cell of the CNT torus. The torus is composed of the six unit cells. }

\end{figure}

\newpage

\begin{figure}
    \epsfxsize=\columnwidth
\centerline{\hbox{
      \epsffile{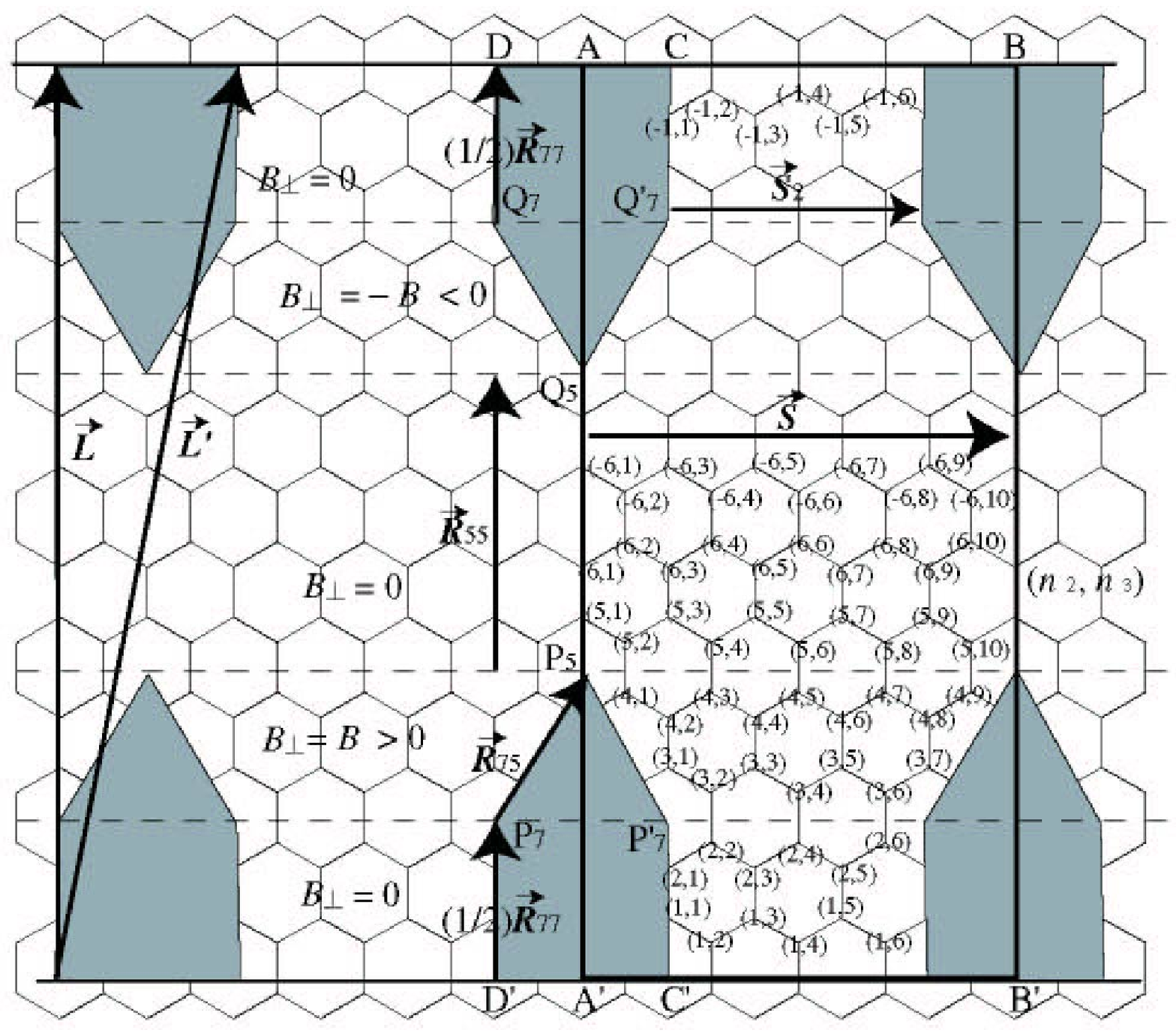}  }}
\caption{
Projection map of the CNT torus. The right rectangle is the unit cell.
Magnetic field perpendicular to this page is denoted by $B_{\bot}$.
 Labels of atoms $(n_2,n_3)$ in the unit cell are shown.
}
\end{figure}

\begin{figure}
    \epsfxsize=0.8\columnwidth
\centerline{\hbox{
      \epsffile{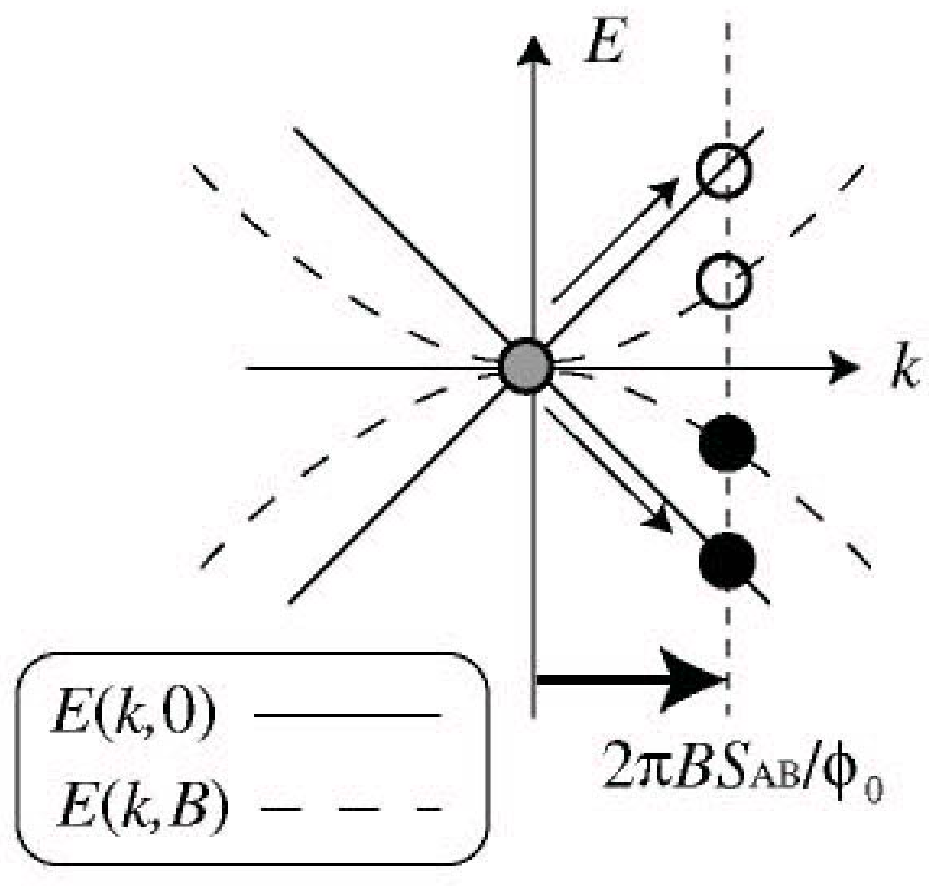}  }}
\caption{
The solid lines and dashed curves 
 represent the dispersion relation
 without and with the direct flux, respectively.
 When the magnetic field $B$
 equals zero, the highest occupied level
 (HOL) of the torus is at the crossing of
 the solid lines as indicated by the gray circle.
 It is degenerate and partially occupied.
 Under finite $B$, the degeneracy is lifted
 and only the lower level (closed circle)
 is occupied while the higher level becomes
 vacant (open circle).
 The arrow indicate the change of the levels
 induced by the AB effect.}
\end{figure}

\begin{figure}
    \epsfxsize=0.9\columnwidth
\centerline{\hbox{
      \epsffile{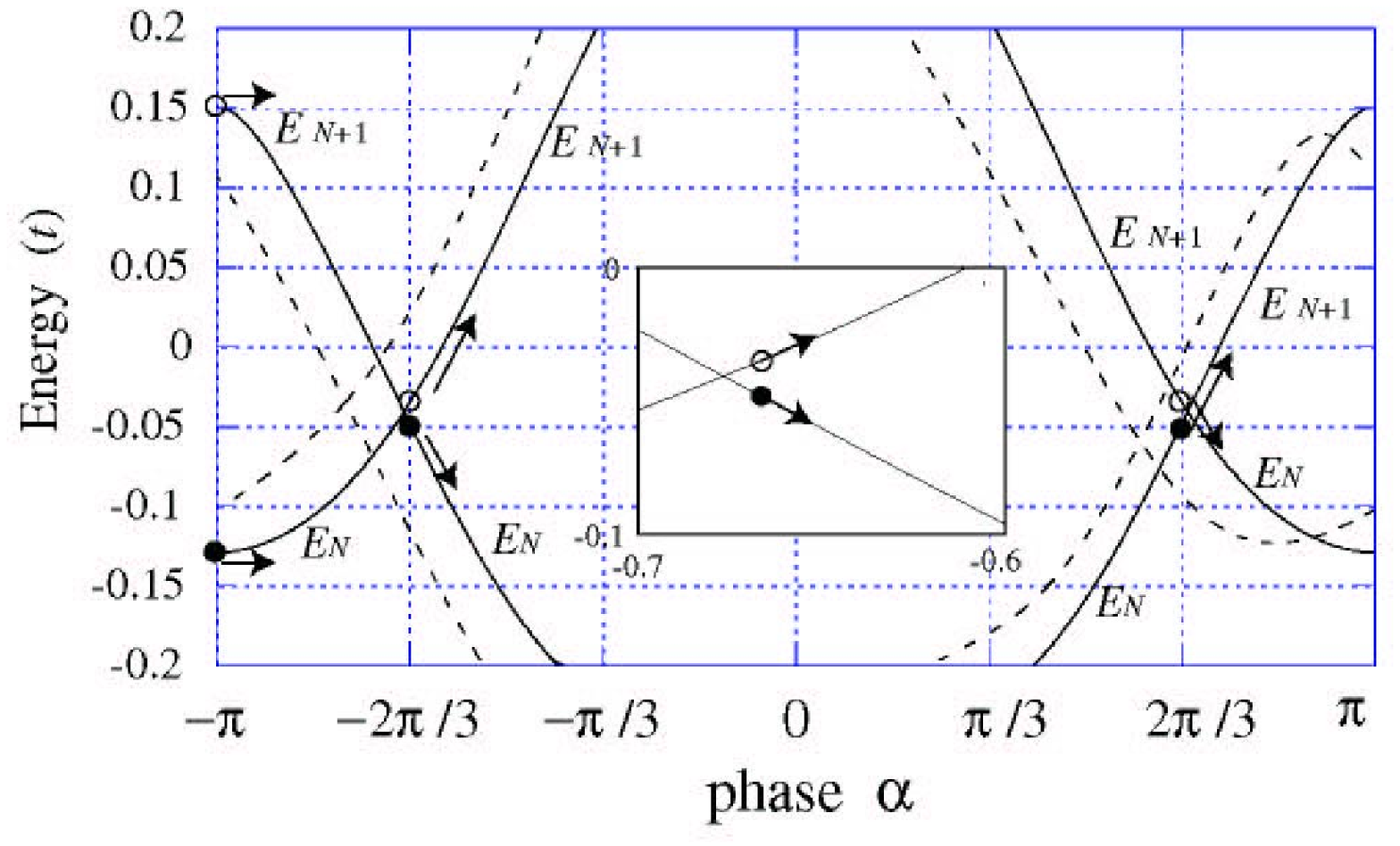}  }}
\caption{
The band structure of the periodic CNT junction of type 2
 with the parameters $(n_S,n_L,n_{75},n_{55})=(5,7,3,2)$.
The circles, lines and arrows
 have the same meanings with Fig.3.
 For the dashed lines,$B=0.01\phi_0/a^2$.
Inset is the
 magnified band structure near the highest occupied level (HOL),
 $E_N(-2\pi/3,0)$.} 
\end{figure}

\begin{figure}
    \epsfxsize=\columnwidth
\centerline{\hbox{
      \epsffile{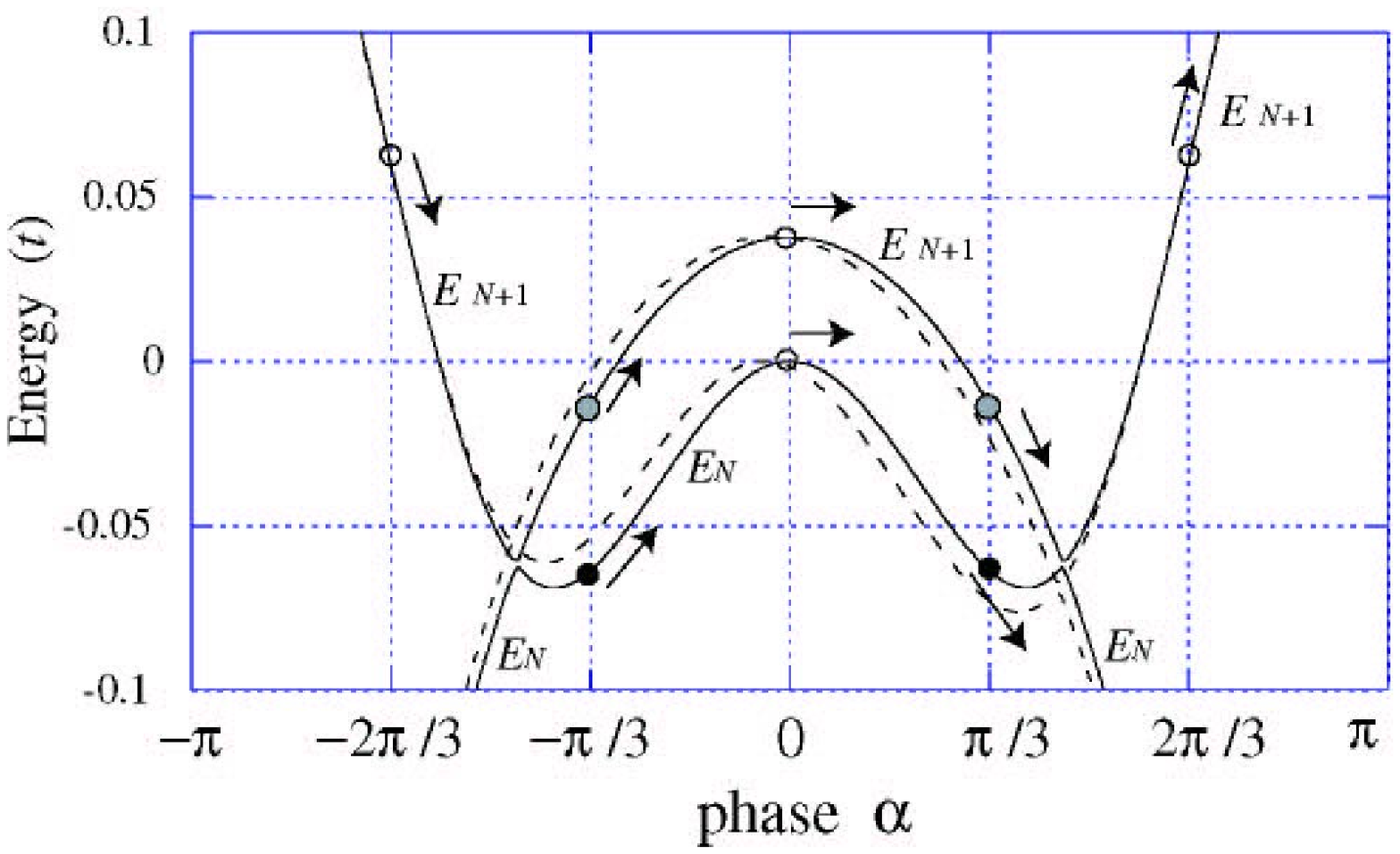}  }}
\caption{
The band structure of the periodic CNT junction of type 3
 with the parameters $(n_S,n_L,n_{75},n_{55})=(3,7,1,3)$.
The circles, lines and arrows
 have the same meanings with Fig.3.
 For the dashed lines,$B=0.01\phi_0/a^2$.
}
\end{figure}

\begin{figure}
    \epsfxsize=\columnwidth
\centerline{\hbox{
      \epsffile{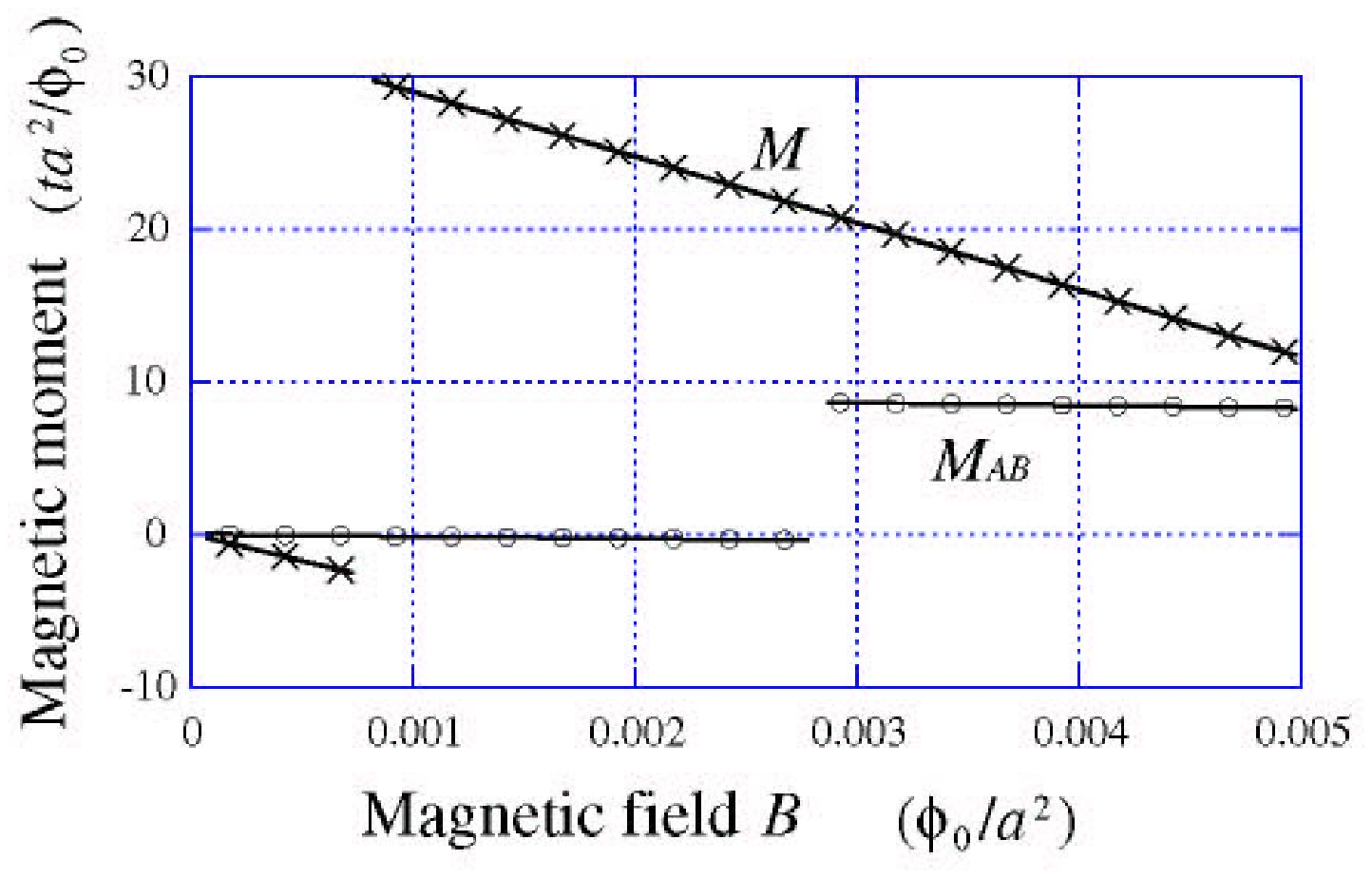}  }}
\caption{
 The magnetic moments 
 of the same periodic CNT junction as that in Fig.4,
 i.e., $(n_S,n_L,n_{75},n_{55})=(5,7,3,2)$,
 as a function of the magnetic field $B$.
 Here 
$M$ and $M_{AB}$ denote the magnetic moment
  induced by full flux and that induced
 by only AB flux, respectively.
 The former is calculated
 by the total energy $U(B)$ as
 $M((j-0.5)\Delta B)=-(U(j\Delta
 B)-U((j-1)\Delta B))/ \Delta B$
 with $\Delta B=0.5 \times 10^{-5} (\phi_0/a^2) $ and
 $j=1,2,\cdots,100$.
 For $M_{AB}$, the total energy is calculated
 by the Hamiltonian where $\gamma^{D}$ in eq.(\protect\ref{alpha})
  is replaced by zero.
}
\end{figure}

\begin{figure}
   \epsfxsize=\columnwidth
\centerline{\hbox{
      \epsffile{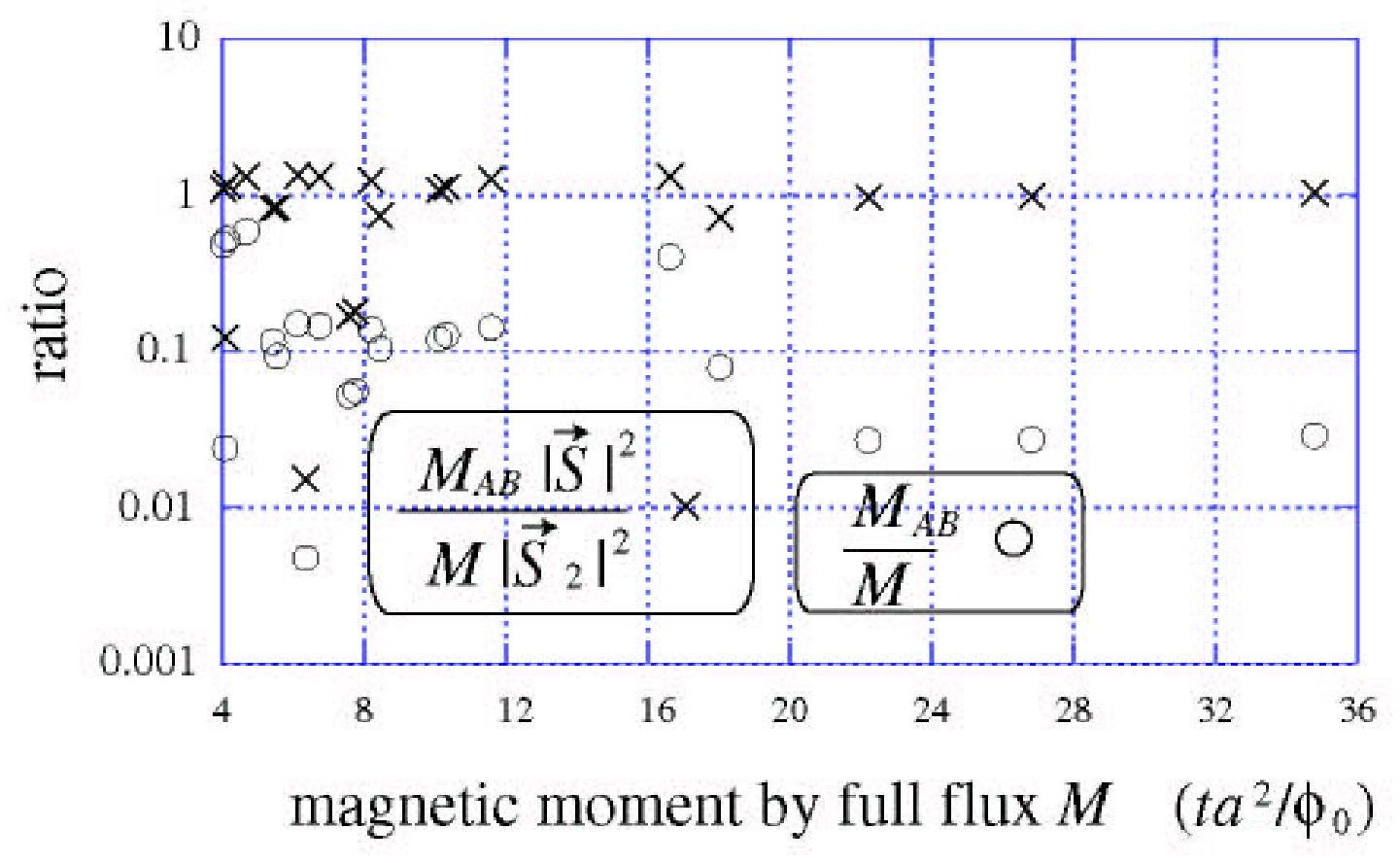}  }}
\caption{
The ratio $M_{AB}/M$ and $(|\vec S|^2/|\vec S_2|^2)M_{AB}/M$ as
 a function of $M$ for the twenty one CNT tori with
 $M$ larger than 4$ta^2/\phi_0$.
 Here $M$ and $M_{AB}$ denote the magnetic moment
  induced by full flux and that induced
 by only AB flux, respectively.
 The applied magnetic field $B$ equals 
 0.75$\times 10^{-5}\phi_0/a^2$.
 The factor $|\vec S|^2/|\vec S_2|^2$ means the ratio
 of the full flux to the AB flux.
 }
\end{figure}


\begin{thebibliography}{10}
\bibitem{note} 
Though we reported that the periodic CNT junctions of type 3
 and type4-1 are semi-metallic in Ref.\cite{helixAkagiprb},
 we found later that they can be also semi-conducting and
 metallic.
\end{thebibliography}
\end{document}